\documentclass[]{emulateapj}

\usepackage{graphicx,times}
\shorttitle{Calibrating overshoot mixing via eclipsing binary stars}
\shortauthors{Meng \& Zhang}
\slugcomment{Submitted 2014 March 23; accepted 2014 April 16 }

\begin{document}

\title{Calibrating the updated overshoot mixing model on eclipsing binary stars: HY Vir, YZ Cas, $\rm{\chi^2}$ Hya \& VV Crv}
\author{Y. Meng\altaffilmark{1,2} \& Q.S. Zhang\altaffilmark{1,3}}
\email{mengy@ynao.ac.cn(YM); zqs@ynao.ac.cn(QSZ)}

\altaffiltext{1}{Yunnan Observatories, Chinese Academy of Sciences, P.O. Box 110, Kunming 650011, China.}
\altaffiltext{3}{University of Chinese Academy of Sciences, Beijing 100039, China.}
\altaffiltext{2}{Key Laboratory for the Structure and Evolution of Celestial Objects, Chinese Academy of Sciences, Kunming, 650011, China.}

\begin{abstract}

The detached eclipsing binary stars with convective cores provide a good tool to investigate the convective core overshoot.
It has been performed on some binary stars to restrict the classical overshoot model which simply extends the boundary of fully mixed region.
However, the classical overshoot model is physically unreasonable and inconsistent with the helioseismic investigation.
An updated model of the overshoot mixing was established recently. There is a key parameter in the model.
In this paper, we use the observations of four eclipsing binary stars, i.e., HY Vir, YZ Cas, $\rm{\chi^2}$ Hya and VV Crv, to investigate the suitable value for the parameter. It is found that the suggested value by the calibrations on eclipsing binary stars is same as the recommended value by other ways.
And we have studied the effects of the updated overshoot model on the stellar structure.
The diffusion coefficient of the convective/overshoot mixing is very high in the convection zone, then quickly decreases near the convective boundary, and exponentially decreases in the overshoot region. The low value of the diffusion coefficient in the overshoot region leads to weak mixing and the partially mixed overshoot region.
The semi-convection, which appears in the standard stellar models of low-mass stars with convective core, is removed by the partial overshoot mixing.

\end{abstract}


\section{Introduction}

The stellar parameters of eclipsing binary stars can be obtained from the analyses of their light curves.
If two components of an eclipsing binary star are detached and the period is long enough, we can assume that each of them obeys the rule of stellar evolutionary theory of a single star.
Those provide the possibility to test the stellar physics in the stellar evolutionary theory.
Especially, the observations of the detached eclipsing binary stars with the masses larger than $\rm{1.2M_{\odot}}$ can be used to restrict the convective core overshoot mixing, which is an important factor to affect the stellar evolution but still not well studied.
That has been performed on some detached eclipsing binary star, e.g., the CO And by \citet{COAnd}, the GX Gem by \citet{GXGem}, and the AQ Ser by \citet{tor14}. \citet{ri00} and \citet{cl07} have studied the dependency of the size of the fully mixed overshoot region on the stellar mass, and have suggested that a classical overshoot region with the size in $0.2\leq \alpha_{OV} \leq 0.25$ is the overall best.

The investigations above are based on the classical overshoot model, which simply extends the convective boundary by a distance to be the boundary of the fully mixed region. That description of the overshoot is based on the 'ballistic' overshoot models (e.g., \citet{sha73,maed75,bre81}). However, the ¡°ballistic¡± overshoot models are excluded by the helioseismic investigation \citep{chr11}. Recently, an updated overshoot mixing model has been established based on fluid dynamics equations \citep{zha13}. The updated model shows that the overshoot mixing can be regarded as a diffusion process, and the efficiency of the mixing in the overshoot region is much lower than that in the convection zone. The key property of this model is the formula of the diffusion coefficient. The formula shows the physical meaning that $\rm{D=C_{OV}L_{Mix}^2/\tau}$ where $\rm{D}$ is the coefficient, $\rm{L_{Mix}}$ is the characteristic length of the overshoot mixing, $\rm{\tau}$ is the characteristic time and $\rm{C_{OV}}$ is a dimensionless parameter which can not be determined by the model itself. The key parameter $\rm{C_{OV}}$ is suggested to be $\rm{10^{-3}}$ \citep{zha13} based on the solar model and the restriction that the equivalent fully mixed overshoot region is less than $\rm{0.4H_P}$. Since the detached eclipsing binary stars provide a good probe on the overshoot mixing, it is necessary to calibrate the key parameter $\rm{C_{OV}}$ by using the observations of detached eclipsing binary stars.

In this paper, we calibrate the parameter $\rm{C_{OV}}$ on four detached eclipsing binary stars with mass ratio being not near unity: HY Vir, YZ Cas, $\rm{\chi^2}$ Hya \& VV Crv, and study the properties of the updated overshoot model. The method of modeling stars and the calibrations are introduced in Section 2. The numerical results of the calibrations and the properties of the updated overshoot model are described in Section 3. The conclusions are summarized in Section 4.

\section{The Method}

\subsection{The stellar evolutionary code and input physics}

The stellar evolutionary code YNEV \citep{zha14a} is adopted to calculate the stellar evolutionary models.
The opacities are interpolated from OPAL opacity tables \citep{OPAL} and the F05 low temperature opacity tables \citep{F05}.
The functions of equation of state (EOS) are interpolated from OPAL-EOS tables \citep{EOS2005}.
The bicubic polynomial is used in the interpolations of opacity and EOS tables in order to obtain continuous derivatives.
The rates of nuclear reactions are based on \citet{Nucl} and enhanced by the weak screening model \citep{sal54}.
The $\rm{T-\tau}$ relation of Eddington gray model of stellar atmosphere is adopted in the atmosphere integral.
In the YNEV code, two theories of stellar convection are optional:
the mixing length theory (MLT) and the turbulent convection model (TCM) developed by \citet{li07}.
The latter is a non-local turbulent convection theory which is based on hydrodynamic equations and some
modeling assumptions. In this paper, we focus on the effects of the overshoot mixing, thus the non-local TCM is adopted to deal with the turbulent convection in stellar interior. The implements of the TCM in the YNEV code are described by \citet{zha12c}.

\subsection{On the overshoot mixing models}

The traditional overshoot mixing model extends the convective boundary by a distance of $\rm{l_{OV}=\alpha_{OV} H_P}$ to be the boundary of the fully mixed region, where $\rm{\alpha_{OV}}$ is a parameter and $\rm{H_P}$ is the pressure scale height. The temperature gradient in this extending region, i.e., the overshoot region, is adiabatic or radiative (radiative means to ignore the convective flux in the overshoot region). The illustration is based on the 'ballistic' overshoot models (e.g., \citet{sha73,maed75,bre81}), which trace the average fluid element overshooting from the convection zone into the radiative region. However, these ¡°ballistic¡± overshoot models are physically unreasonable \citep{ren87,zha13} and inconsistent with the helioseismic investigations. The helioseismic investigations \citep{chr11} have shown that the temperature gradient smoothly changes from adiabatic to radiative near the convective boundary, so the 'ballistic' overshoot models are excluded because they show adiabatic overshoot region and a jump of the temperature gradient at the boundary of the overshoot region. It has been suggested that only the turbulent convection models (e.g., \citet{xio81,xio97,can97,can98,den06,li07,can11,li12}) can fit the restriction.

Another popular overshoot mixing model is the diffusion model with the diffusion coefficient $\rm{D}$ based on the characteristic turbulent velocity $\rm{v}$ and the characteristic length $\rm{l}$, i.e., $\rm{D \propto vl}$. However, in most diffusion overshoot mixing models (e.g., \citet{fre96,ven98,lai11,zhali12,ding14}), the characteristic length $\rm{l}$ is assumed to be comparable with $\rm{H_P}$. This is an analogy to assume the characteristic length in the overshoot region being similar with the characteristic length in the convection zone. \citet{den96} have found that setting $\rm{l \sim H_P}$ in $\rm{D \sim  vl/3}$ leads to almost fully mixing and have suggested a small characteristic length as $\rm{ l \sim 10^{-5} l_0 \sim 10^{-5} H_P }$ for the overshoot mixing in order to result in a mixing time scale being comparable with the evolutionary time scale. \citet{zha12a,zha12b,zha12c} have shown that when the characteristic length is assume to be comparable with $\rm{H_P}$, the dimensionless parameter in $\rm{D = C_X v H_P}$ should be small as $\rm{C_X \sim 10^{-10}}$ in order to fit some observations. This excessively small dimensionless parameter makes the assumption $\rm{l \sim H_P}$ be doubtful.

Recently, \citet{zha13} has developed an updated overshoot mixing model based on hydrodynamic equations and some
modeling assumptions.
This model focuses on the turbulent flux of the chemical component and calculates the diffusion coefficient for convective/overshoot mixing.
It is found in the model that the diffusion coefficient in overshoot region is different from that in convection zone.
In the convection zone,
\begin{eqnarray} \label{difcz}
D_{CZ} = C_{CZ} \frac{k^2}{\varepsilon},
\end{eqnarray}%
and in the overshoot region,
\begin{eqnarray} \label{difov}
D_{OV} = C_{OV} \frac{\varepsilon}{N^2},
\end{eqnarray}%
where $\rm{k}$ is the turbulent kinetic energy, $\rm{\varepsilon}$ is the turbulent dissipation rate, $\rm{N^2}$ describes Brunt-V\"{a}is\"{a}l\"{a} frequency, $\rm{C_{CZ}}$ is a parameter of the magnitude order of unity, and $\rm{C_{OV}}$ is another parameter which is recommended to be on the magnitude order $\rm{C_{OV}}\sim 10^{-3}$ based on the adopted TCM (and its parameters) and some observational restrictions \citep{zha13}.
The physical meanings of Equation (1) and (2) have also been pointed out. Equation (1) is equivalent to the model $\rm{D \propto vH_P}$. However, Equation (2) is physically different from $\rm{D \propto vH_P}$. The diffusion coefficient in the overshoot region being Equation (2) is for the reason that fluid elements moving around their equilibrium location so the characteristic length is $\rm{v/N}$ \citep{zha13}.

In this paper, we use Zhang's (2013) overshoot mixing model (i.e., Equation (2)). The turbulent dissipation rate $\rm{\varepsilon}$ is calculated by using the TCM \citep{li07}. As the value of the dimensionless parameter $\rm{C_{OV}}$ for overshoot mixing is not determined in the theoretical model, the main aim of this paper is to use the observations of eclipsing binary stars to restrict the value of $\rm{C_{OV}}$.

\subsection{On calibrating overshoot mixing via eclipsing binary stars}

The masses and radii of two components of an eclipsing binary star (we mean detached eclipsing binary star in this paper) can be observed via the analyses of light curves. The masses, radii and the effective temperatures can be used to restrict the stellar evolution theory: the evolutionary track of a star with given mass should pass the observed radius and effective temperature. For a main-sequence star with a convective core, the evolutionary track is sensitive to the convective overshoot. Therefore, the eclipsing binary is a good tool to investigate the convective overshoot. This has been performed on some eclipsing binary stars, e.g., the CO And by \citet{COAnd}, and the AQ Ser by \citet{tor14}. \citet{ri00} and \citet{cl07} have studied the dependency of the size of the fully mixed overshoot region to the stellar mass, suggested that a classical overshoot region with the size in $0.2\leq \alpha_{OV} \leq 0.25$ is the overall best.

In the investigations mentioned above, the overshoot region is assumed to be fully mixed. It is necessary to study the updated overshoot formula via the eclipsing binary stars.

In the cases of standard stellar models with convective core, the structure of a star is fixed when the mass, initial hydrogen abundance $\rm{X}$, initial metallicity $\rm{Z}$, the age $\rm{t}$, an overshoot parameter (i.e.., $\rm{\alpha_{OV}}$ for the classical overshoot or $\rm{C_{OV}}$ for the updated overshoot model) and a convection parameter $\rm{\alpha}$ (i.e., $\rm{\alpha_{MLT}}$ for the MLT theory or $\rm{\alpha_{TCM}}$ for the TCM theory) are all fixed. The observations of a binary star give four restrictions for two stars with given masses, i.e., the radii and effective temperatures of two components.
We assume that the age and the chemical composition are same for two components.
In this case, the initial hydrogen abundance $\rm{X}$, initial metallicity $\rm{Z}$, the age $\rm{t}$ and the overshoot parameter of the binary star can be mathematically fixed when we adopt a fixed convection parameter because the number of variables is equal to the number of the equations (Equations (\ref{Appx-A1}) for two components). The standard errors can be also obtained based on the method described in the Appendix. According to those properties, \citet{zha12c} has tested the previous diffusion formula of the overshoot on the binary star HY Vir.

However, this method of calibrating does not work in some cases. When the mass ratio $\rm{q\approx1}$, the masses, radii and effective temperatures of two components are very close to each other. This leads to the problem that the Equations (\ref{Appx-A1}) for the primary are almost identical to Equations (\ref{Appx-A1}) for the secondary, so there are only two independent equations. Therefore, we can do the calibration only on the eclipsing binary stars with the masses of two components being obviously different. The effects of the overshoot mixing on the stellar radius and effective temperature are accumulating as the stellar age increasing. Therefore, we can not calibrate the overshoot parameter by using the stars near the ZAMS stage (e.g., UZ Dra \citep{UZDra} and V335 Ser \citep{V335Ser}).

The observation data for intermediate- and high-mass eclipsing binaries are few and not accurate enough. For the high-mass stars, the mass loss \citep{chio78,bru82,chio86,maed87,mey94} and rotation \citep{mey00,bro11,maed12} can also significantly affect the stellar structure and evolution. Since the rotation and mass loss are not well studied at present, we do not attempt to calibrate the overshoot in high-mass stars. We focus on the low-mass eclipsing binaries.
In this paper, we use the methods above to find the suitable value of the overshoot parameter $\rm{C_{OV}}$ of the updated overshoot mixing model on four eclipsing binaries: HY Vir, YZ Cas, $\rm{\chi^2}$ Hya \& VV Crv.

\section{Numerical results}

In this section, we show the results of the calibration on the eclipsing binaries and properties of the updated overshoot mixing model in stellar interior. All of the stellar models evolve from the pre-main sequences with the center temperature $\rm{T_C=10^5K}$.
The metal composition is assumed to be same as the solar metal composition AGSS09 \citep{AGSS09}.
The TCM \citep{li07} is adopted to calculate the turbulent variables (e.g., turbulent dissipation rate required in the overshoot diffusion coefficient and the convective flux). The parameters of the TCM are same as \citet{zha12c}. Turbulent dissipation parameter $\rm{\alpha_{TCM}=0.8}$ is based on solar calibration with the AGSS09 composition. The number of mesh points in stellar models is typically a thousand.

\subsection{Calibration on eclipsing binary models}

We solve equation (\ref{Appx-A1}) for two components to calibrate the overshoot parameter $\rm{C_{OV}}$, composition ($\rm{X}$ and $\rm{Z}$) and age $\rm{t}$ on four eclipsing binaries: HY Vir, YZ Cas, $\rm{\chi^2}$ Hya and VV Crv. The mass range in the samples is about $\rm{1.3<M/M_{\odot}<3.6}$ which comprises seven low-mass stars and an intermediate-mass star (i.e., the primary of $\rm{\chi^2}$ Hya). The results of the calibrations are shown in Table 1. The radii and the effective temperatures of the calibrated stellar models match the observations in the accuracies of $\rm{\Delta R/R_{\odot} <10^{-3}}$ and $\rm{\Delta lgT_{eff}<10^{-3}}$.

\begin{table*}
\centering
\caption{ Parameters of the eclipsing binaries: HY Vir, YZ Cas, $\rm{\chi^2}$ Hya \& VV Crv. } \label{Tab1}
\begin{tabular}{lcccc}
\hline\hline
 & HY Vir & YZ Cas & $\rm{\chi^2}$ Hya & VV Crv  \\\hline
 Observations & & & & \\\hline
$\rm{M_A/M_{\odot}}$  & $\rm{1.838\pm0.009}$ & $\rm{2.263\pm0.012}$ & $\rm{3.613\pm0.079}$ & $\rm{1.978\pm0.010}$  \\
$\rm{M_B/M_{\odot}}$  & $\rm{1.404\pm0.006}$ & $\rm{1.325\pm0.007}$ & $\rm{2.638\pm0.050}$ & $\rm{1.513\pm0.008}$  \\
$\rm{R_A/R_{\odot}}$  & $\rm{2.806\pm0.008}$ & $\rm{2.525\pm0.011}$ & $\rm{4.384\pm0.039}$ & $\rm{3.375\pm0.010}$  \\
$\rm{R_B/R_{\odot}}$  & $\rm{1.519\pm0.008}$ & $\rm{1.331\pm0.006}$ & $\rm{2.165\pm0.043}$ & $\rm{1.650\pm0.008}$  \\
$\rm{lg(T_A)}      $  & $\rm{3.836\pm0.008}$ & $\rm{3.979\pm0.006}$ & $\rm{4.066\pm0.010}$ & $\rm{3.813\pm0.013}$  \\
$\rm{lg(T_B)}      $  & $\rm{3.816\pm0.008}$ & $\rm{3.838\pm0.015}$ & $\rm{4.041\pm0.010}$ & $\rm{3.822\pm0.013}$  \\
Ref. $\rm{Z}$  & $\rm{0.027}$ & $\rm{0.009\pm0.003}$ & $\rm{0.025\pm0.010}$ & $\rm{0.034\pm0.013}$  \\
Ref. $\rm{t/Gyr}$  & $\rm{1.35\pm0.10}$ & $\rm{0.52\pm0.03}$ & $\rm{0.18\pm0.02}$ & $\rm{1.2\pm0.1}$  \\
 Ref. &  (1)  & (2) & (3) & (4) \\\hline
 Calibrations & & & & \\\hline
$\rm{C_{OV}\times10^{3}}$  & $\rm{0.9\pm0.4}$ & $\rm{0.9\pm0.4}$ & $\rm{0.7\pm3.0}$ & $\rm{0.4\pm1.5}$ \\
$\rm{X}            $  & $\rm{0.64\pm0.03}$ & $\rm{0.71\pm0.02}$ & $\rm{0.72\pm0.05}$ & $\rm{0.7\pm0.1}$  \\
$\rm{Z}            $  & $\rm{0.032\pm0.009}$ & $\rm{0.012\pm0.002}$ & $\rm{0.013\pm0.003}$ & $\rm{0.03\pm0.01}$  \\
$\rm{t/Gyr}        $  & $\rm{1.3\pm0.1}$ & $\rm{0.55\pm0.07}$ & $\rm{0.21\pm0.09}$ & $\rm{1.3\pm0.4}$  \\
\hline\hline
\end{tabular}
\tablecomments{References: (1) \citet{HYVir}; (2) \citet{YZCas}; (3) \citet{X2Hya}; (4) \citet{VVCrv}. }
\end{table*}

The calibration results in Table 1 show that the best value of $\rm{C_{OV}}$ is about $\rm{1\times10^{-3}}$. This is consistent with the suggested value via the test of solar model and the classical restriction on convective core overshoot \citep{zha13}.
Although the mass of two components are different and there may be possible relationship between $\rm{C_{OV}}$ and stellar mass, we fix $\rm{C_{OV}}$ in the calibration of each eclipsing binary. However, the results do not support an obvious dependency of $\rm{C_{OV}}$ on stellar mass.
The calibration results show that the standard errors of $\rm{C_{OV}}$ are significant. It seems that the observational errors of $\rm{M/M_{\odot}}$, $\rm{R/R_{\odot}}$ and especially $\rm{lgT_{eff}}$ should be less than $\rm{1\%}$, otherwise the corresponding error of the overshoot parameter is too large.
The metallicities and ages of those eclipsing binaries have been suggested in the references as shown in Table 1 (e.g., Ref. Z and t) by comparing the stellar model based on fully mixed classical overshoot region with the observations. It is found that our results of metallicities and ages based on the updated overshoot mixing model are similar to the results of classical overshoot model. Only the suggested metallicity of $\rm{\chi^2}$ Hya is significantly higher than our calibration. This may result from that \citet{X2Hya} have used old opacity tables in modeling stars. The calibrations of the chemical composition on the HY Vir and YZ Cas support a helium enrichment law $\rm{\Delta Y / \Delta Z \approx 2}$. The results of the chemical composition of $\rm{\chi^2}$ Hya and VV Crv are not accurate enough to validate the law.

\begin{figure}
\plotone{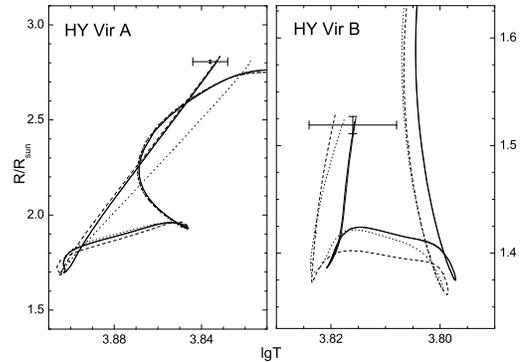}
\caption{The evolutionary tracks on radius vs. effective temperature for HY Vir. The thick solid lines are for the model of $\rm{C_{OV}=1\times10^{-3}}$, the solid lines are for the calibrated stellar model, the dashed lines are for the stellar model with $\rm{0.3H_P}$ fully mixed classical overshoot region, and the dotted lines are for the standard stellar models without mixing outside the convection zones. The left plane is for the primary of HY Vir, and the right plane is for the secondary. } \label{HYVir}
\end{figure}

\begin{figure}
\plotone{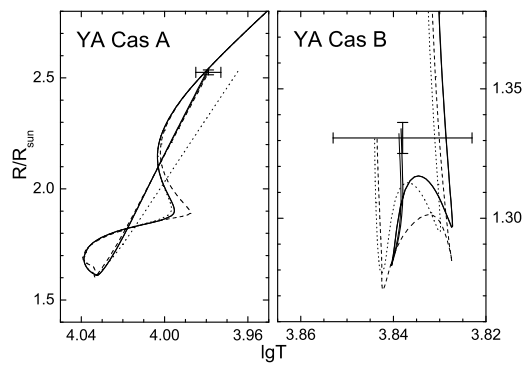}
\caption{Similar to Fig.(\ref{HYVir}), but for YZ Cas. } \label{YZCas}
\end{figure}

\begin{figure}
\plotone{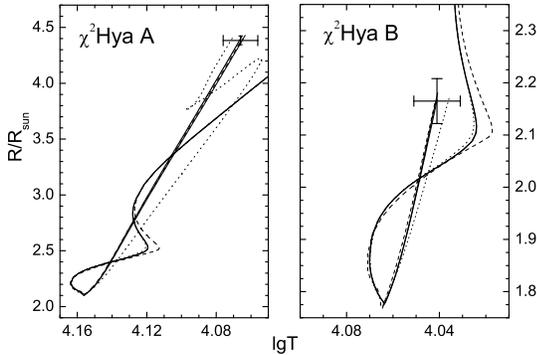}
\caption{Similar to Fig.(\ref{HYVir}), but for $\rm{\chi^2}$ Hya. } \label{X2Hya}
\end{figure}

\begin{figure}
\plotone{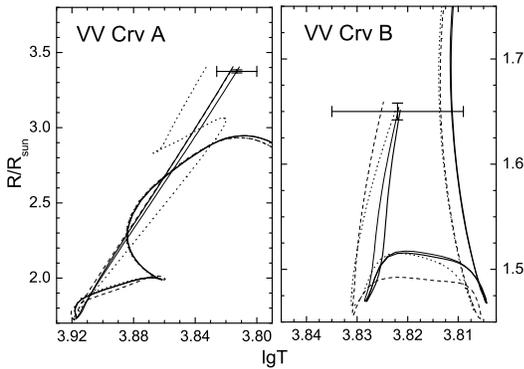}
\caption{Similar to Fig.(\ref{HYVir}), but for VV Crv. } \label{VVCrv}
\end{figure}

The evolutionary tracks on effective temperature vs. radius for the stellar models of the binaries are shown in Figs.(\ref{HYVir}-\ref{VVCrv}).
The thick solid lines, the solid lines, the dashed lines and the dotted lines are for the evolutionary tracks with $\rm{C_{OV}=1\times10^{-3}}$, the calibrated stellar models, the stellar models with $\rm{0.3H_P}$ fully mixed classical overshoot region and the standard stellar models without mixing outside the convection zones, respectively.
The stellar models with $\rm{0.3H_P}$ fully mixed overshoot region and the standard stellar models without mixing are calculated for comparison, and the mixing length theory (MLT) is adopted in those models to deal with the convective flux. The parameter $\rm{\alpha_{MLT}=1.75}$ in the MLT is based on the solar calibration.
The dashed lines are almost identical with the thick solid lines, indicating that the updated overshoot model with $\rm{C_{OV}=1\times10^{-3}}$ leads to the similar mixing efficiency as a $\rm{0.3H_P}$ fully mixed overshoot region in the stars with the mass in the studied range.
It is shown that the stellar effective temperature of the models in the PMS stage and near the ZAMS is slightly affected by the adopted convection theory (MLT is used in dashed lines and dotted lines, and the non-local TCM is used in solid lines and thick solid lines). But the differences are small because both the turbulent dissipation parameters in the MLT and the TCM are based on solar calibrations and both convection theories show adiabatic convection in the convective core.
It is interesting that, for the stars with $\rm{M/M_{\odot}>2}$ (i.e., twos components of $\rm{\chi^2}$ Hya and the primary of YZ Cas), there are significant differences in the PMS stage, i.e., where $\rm{R/R_{\odot}\approx2.5}$ for the primary of $\rm{\chi^2}$ Hya, $\rm{R/R_{\odot}\approx2.1}$ for the secondary of $\rm{\chi^2}$ Hya, and $\rm{R/R_{\odot}\approx1.9}$ for the primary of YZ Cas. This can be explained as follows. At those locations, $\rm{^{12}C}$ is burned to be $\rm{^{14}N}$ in the center. The overshoot mixing could affect this process because of the refueling of $\rm{^{12}C}$ in the convective core. In the stellar models with $\rm{0.3H_P}$ fully mixed overshoot region, the overshoot mixing is assumed as instantaneous mixing. In the updated overshoot model, the overshoot is a diffusion process, and the diffusion coefficient is not high enough to result in significant mixing in the short time scale of the PMS stage. Therefore, in Fig.(\ref{X2Hya}) and the left plane in Fig.(\ref{YZCas}), the solid lines and the thick solid lines locate between dotted lines and dashed lines, and are very close to the dotted lines.

\subsection{Properties of the updated overshoot mixing model in the core overshoot region of low-mass stars}

\begin{figure*}
\centering
\plotone{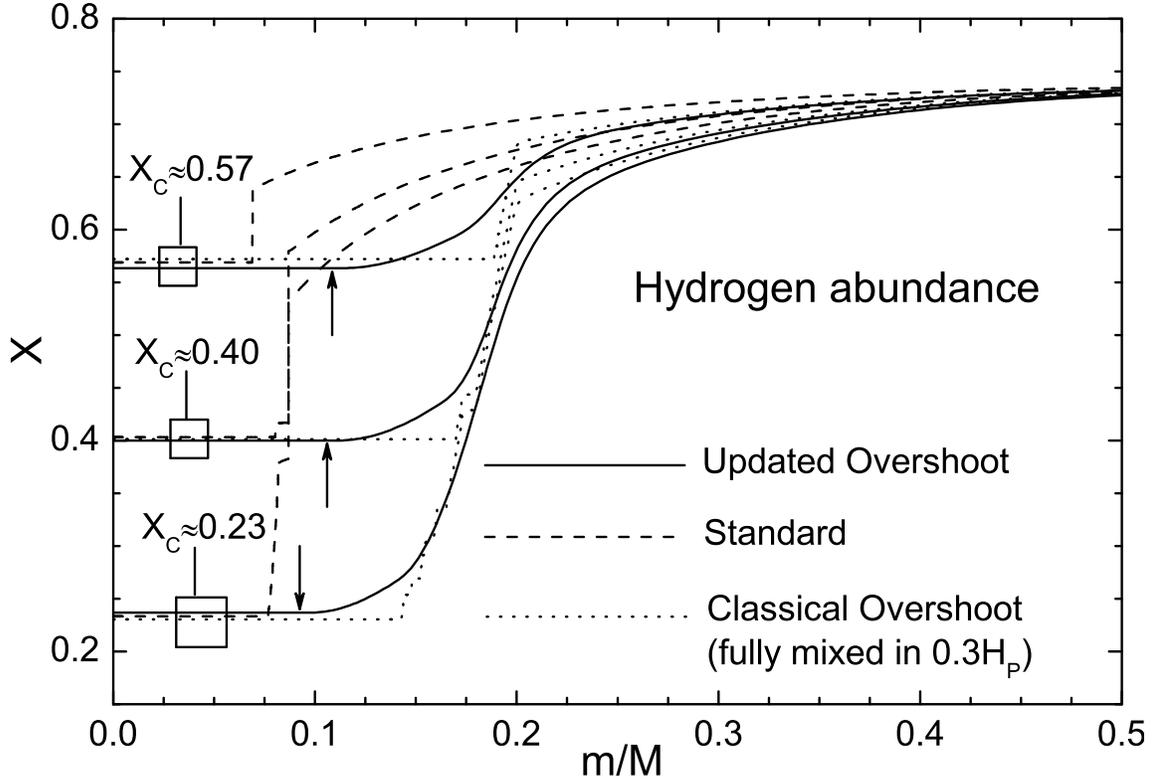}
\caption{The hydrogen abundance in stellar interior. The arrows indicate the convective boundary of the models for the solid lines. } \label{XProf}
\end{figure*}

The hydrogen abundance in the stellar interior models are shown in Fig.\ref{XProf}. Models in three cases, i.e., the standard model, the model with classical overshoot with $\rm{0.3H_P}$ and the updated overshoot model with $\rm{C_{OV}=10^{-3}}$, and with different center hydrogen abundance ($\rm{X_C\approx0.57}$, $\rm{0.40}$, $\rm{0.23}$) are shown.
It is found that the updated overshoot model results in smooth profile of the hydrogen abundance. Near the convective boundary, the gradient is close to zero, indicating large diffusion coefficient. Comparing with the standard stellar models and the classical overshoot stellar models, the stellar models with updated overshoot model shows similar effects to the classical overshoot model with $\rm{0.3H_P}$ on refueling the core. This can be validated by estimating the difference of the areas below the solid lines and corresponding dotted lines. The distinction is that, the hydrogen abundances of the updated overshoot model are always smooth, unlike that there is no derivatives at the fully mixing boundaries in the classical overshoot model. The updated overshoot model showing similar effects as the classical overshoot model with $\rm{0.3H_P}$ explains that, as mentioned above, the evolutionary tracks of the two cases are identical and the calibrated metallicity and age are consistent with the suggested values based on stellar models with classical overshoot.

\begin{figure}
\plotone{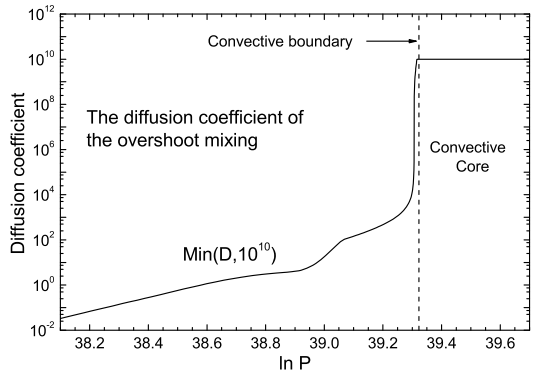}
\caption{The diffusion coefficient of the convective overshoot mixing. $\rm{D}$ is the diffusion coefficient in the unit of $\rm{cm^2 s^{-1}}$. $\rm{P}$ is the pressure in stellar interior in the unit of $\rm{g cm^{-1}s^{-2}}$.} \label{DIF}
\end{figure}

The formula of the diffusion coefficient of the updated overshoot mixing model (e.g., Equation (26) in \citet{zha13}) has shown that the diffusion coefficient is very large in the convection zone and is low in the overshoot region, which in intrinsic ensures a fully mixed convection zone and a partially mixed overshoot region \citep{zha13}. The approximation of the formula of the diffusion coefficient in the overshoot region is Equation (\ref{difov}), which is adopted in the numerical calculations. Figure \ref{DIF} shows the profile of the diffusion coefficient for the overshoot mixing. The stellar model is for the $\rm{1.5M_{\odot}}$ star with $\rm{X_C=0.4}$ and $\rm{C_{OV}=10^{-3}}$. According to Equation \ref{difcz}, the diffusion coefficient in the convection zone is as large as $\rm{D\sim 10^{16}}$. We set an upper limit of $\rm{D}$ as $\rm{10^{10}}$ in the calculations because excessively large diffusion coefficient may lead to numerical instability and $\rm{D>10^{10}}$ ensures complete mixing. It is shown that $\rm{D}$ quickly decreases in a thin layer near the convective boundary and then exponentially decreases in the most part of overshoot region. After the quick decreasing, the geometric mean diffusion coefficient in the overshoot region is typically $\rm{10^2}$ and $\rm{10^0}$ for $\rm{0.5H_P}$ and $\rm{1H_P}$, respectively. Accordingly, the time scale for a length $\rm{L}$ showing obvious mixing is $\rm{\tau \sim L^2/D}$, i.e, about $\rm{10^{16}s}$ for $\rm{0.5H_P}$ and about $\rm{10^{20}s}$ for $\rm{1H_P}$ ($\rm{H_P \approx 10^{10}cm}$ here). The former is on the same magnitude order as the evolutionary time scale and the latter is much larger than the evolutionary time scale.

\begin{figure*}
\centering
\plottwo{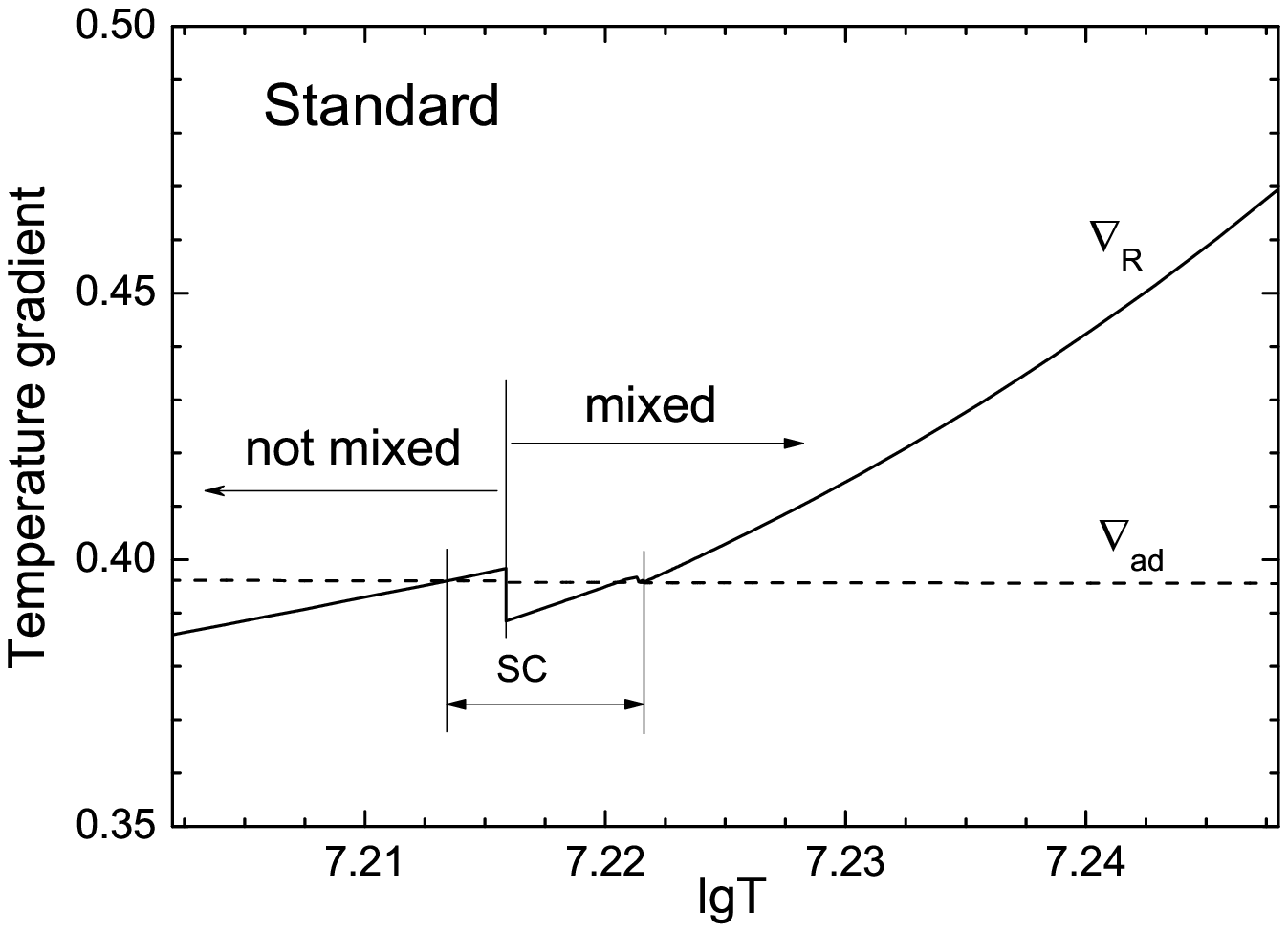}{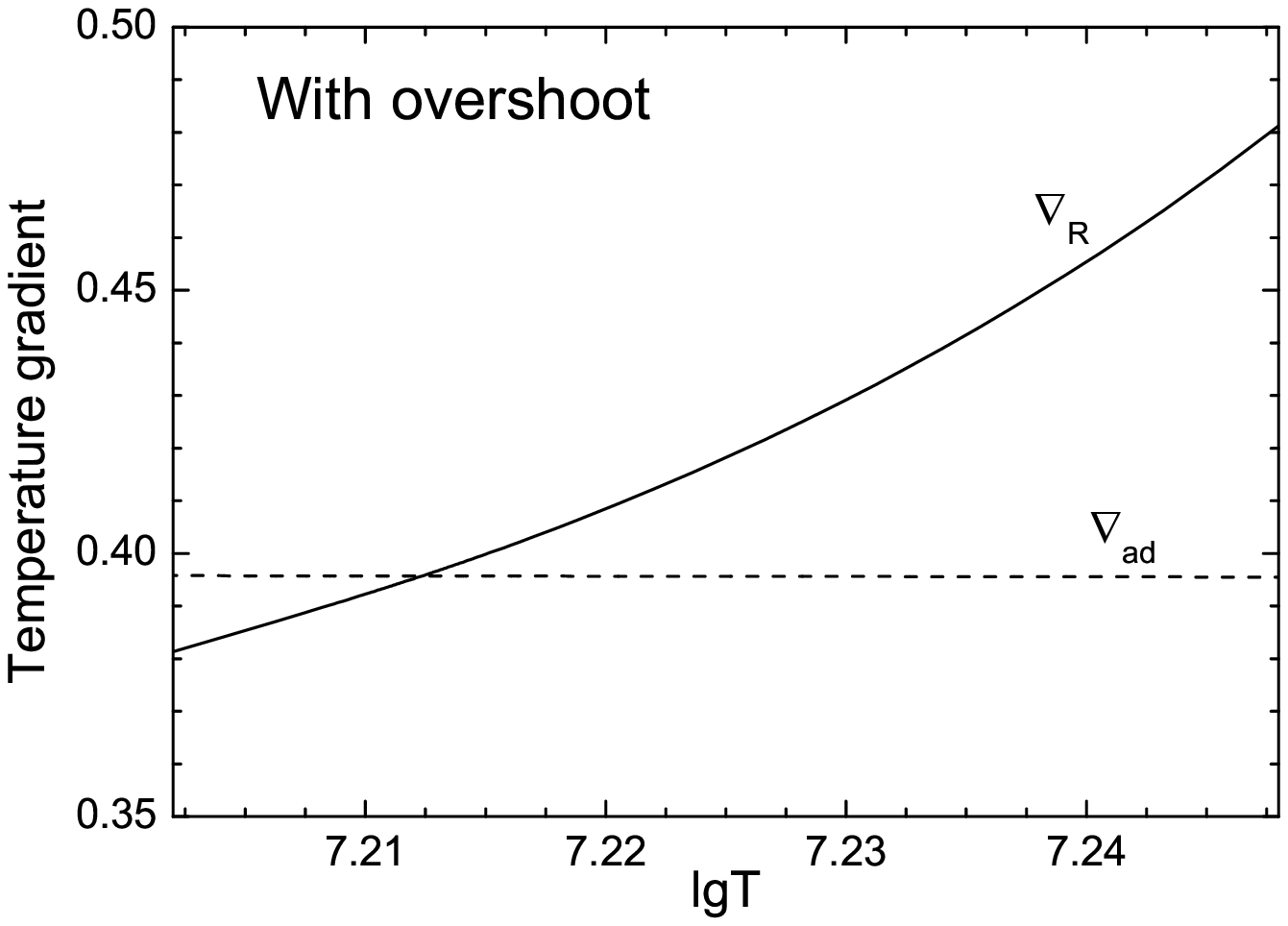}
\caption{The convection status near the convective boundary for the standard stellar model and the stellar model with updated overshoot model. $\rm{\nabla_R}$ and $\rm{\nabla_{ad}}$ are the dimensionless radiative temperature gradient and the adiabatic radiative temperature gradient, respectively.} \label{SC}
\end{figure*}

In the standard stellar model of a low-mass main sequence star with convective core, a phenomenon called the semi-convection occurs. There is a region near the convective boundary, $\rm{\nabla_R > \nabla_{ad}}$ when this region is not mixed into the convective core, or $\rm{\nabla_R < \nabla_{ad}}$ when this region is fully mixed into the convective core. This leads to the contradiction that the fully mixed region is not the convection boundary. In the framework of classical ideal of the local convection, this region should be partially mixed to reach the convective neutral condition $\rm{\nabla_R = \nabla_{ad}}$, since the mixing process can not continue when the neutral condition is satisfied \citep{sch58}. The intrinsic reason of this phenomenon is that the mixing time scale is much less than the evolutionary time scale (the mixing time scale is zero in the classical ideal local convection, e.g., instantaneous mixing) thus we do not have enough time resolution to trace the variation of $\rm{\nabla_R}$ during the mixing process. However, the semi-convection does not appear in our stellar models with the updated overshoot mixing model. Figure \ref{SC} shows the profiles of $\rm{\nabla_R}$ and $\rm{\nabla_{ad}}$ in stellar models with $\rm{M=1.5M_{\odot}}$ and $\rm{X_C \approx 0.4}$. The left plot shows the standard model and the right plot shows the model with updated overshoot mixing. In the region denoted as 'SC', it is clearly shown that $\rm{\nabla_R > \nabla_{ad}}$ when the region is not mixed and $\rm{\nabla_R < \nabla_{ad}}$ when the region is fully mixed into the convective core. The radiative temperature gradient $\rm{\nabla_R}$ is discontinuous at the fully mixing boundary due to the jump of the chemical abundance. In the model with updated overshoot, there is no such phenomenon. The radiative temperature gradient is continuous because the updated overshoot model describes a weak mixing process and the profile of chemical abundance is continuous. \citet{xio81,xio86} has found that the semi-convection in massive stars results from the local convection theory and can be removed by using the non-local turbulent convection theory. Our calculations show the similar result that the non-local effect of the turbulent convection, i.e., the overshoot mixing, can remove the semi-convection in low-mass stars.

\section{Conclusions}

In this paper, we have used the observations of four eclipsing binary stars, i.e., HY Vir, YZ Cas, $\rm{\chi^2}$ Hya and VV Crv, to calibrate the updated overshoot mixing model recently developed by \citet{zha13}. And we have investigated the basic properties of stellar structures based on the model. The main results are as follows.

The dimensionless parameter in the updated overshoot mixing model is suggested to be $\rm{C_{OV}=10^{-3}}$ for low mass stars. Stellar models with this value can fit the observations of the concerned eclipsing binary stars in $\rm{1 \delta }$. No obvious dependency of $\rm{C_{OV}}$ on the stellar mass is found in the low-mass stars with $\rm{1.2<M/M_{\odot}<2.5}$. The suggested value of $\rm{C_{OV}}$ in this paper is same to \citet{zha13}, but using a different method.

The updated formula of the overshoot mixing shows that the diffusion coefficient quickly decreases near the convective boundary and exponentially decreases in the most part of the overshoot region. This leads to a partial mixing region out side the convective core. The efficiency of the overshoot mixing is high in the thin layer near the convective boundary due to the high diffusion coefficient, but is low in most of the overshoot region. The semi-convection, which appears in the standard stellar models of low-mass star with convective core, is removed by the partial overshoot mixing.

\acknowledgments

Many thanks to the anonymous referee for careful reading of the manuscript and providing comments which improved the original version.
This work is co-sponsored by the National Natural Science Foundation of China (NSFC) through grant No. 11303087, the Science Foundation of Yunnan Observatory No. Y1ZX011007 \& Y3CZ051005, the West Light Foundation of the Chinese Academy of Sciences and the Chinese Academy of Sciences under grant No. KJCX2-YW-T24.

\appendix
\section{The Calculations of the Standard Errors}

In this appendix, we show the details of obtaining the standard errors of ($\rm{C_{OV}}$, $\rm{X}$, $\rm{Z}$, $\rm{t}$) based on the standard errors of observed ($\rm{lgT_{A}}$, $\rm{lgT_{B}}$, $\rm{R_{A}}$, $\rm{R_{B}}$, $\rm{M_{A}}$, $\rm{M_{B}}$).

The effective temperature and the radius of a star is determined by the mass $\rm{M}$, overshoot parameter $\rm{C_{OV}}$, hydrogen abundance $\rm{X}$, metallicity $\rm{Z}$ and its age $\rm{t}$, i.e.,
\begin{eqnarray} \label{Appx-A1}
\begin{array}{l}
\begin{array}{l}
 T = T(M,{C_{OV}},X,Z,t), \\
 R = R(M,{C_{OV}},X,Z,t). \\
 \end{array}
 \end{array}
\end{eqnarray}%
Accordingly, we have the relation between the variations of ($\rm{C_{OV}}$, $\rm{X}$, $\rm{Z}$, $\rm{t}$) and the variations of ($\rm{lgT_{A}}$, $\rm{lgT_{B}}$, $\rm{R_{A}}$, $\rm{R_{B}}$, $\rm{M_{A}}$, $\rm{M_{B}}$) as follows:
\begin{eqnarray} \label{Appx-relationD}
\begin{array}{l}
 \left[ {\begin{array}{*{20}{c}}
   {d{C_{OV}}}  \\
   {dX}  \\
   {dZ}  \\
   {dt}  \\
\end{array}} \right] = {\left[ {\begin{array}{*{20}{c}}
   {\frac{{\partial \lg {T_A}}}{{\partial {C_{OV}}}}} & {\frac{{\partial \lg {T_A}}}{{\partial X}}} & {\frac{{\partial \lg {T_A}}}{{\partial Z}}} & {\frac{{\partial \lg {T_A}}}{{\partial t}}}  \\
   {\frac{{\partial \lg {T_B}}}{{\partial {C_{OV}}}}} & {\frac{{\partial \lg {T_B}}}{{\partial X}}} & {\frac{{\partial \lg {T_B}}}{{\partial Z}}} & {\frac{{\partial \lg {T_B}}}{{\partial t}}}  \\
   {\frac{{\partial {R_A}}}{{\partial {C_{OV}}}}} & {\frac{{\partial {R_A}}}{{\partial X}}} & {\frac{{\partial {R_A}}}{{\partial Z}}} & {\frac{{\partial {R_A}}}{{\partial t}}}  \\
   {\frac{{\partial {R_B}}}{{\partial {C_{OV}}}}} & {\frac{{\partial {R_B}}}{{\partial X}}} & {\frac{{\partial {R_B}}}{{\partial Z}}} & {\frac{{\partial {R_B}}}{{\partial t}}}  \\
\end{array}} \right]^{ - 1}} \\
  \times \left[ {\begin{array}{*{20}{c}}
   {d\lg {T_A} - \frac{{\partial \lg {T_A}}}{{\partial {M_A}}}d{M_A}}  \\
   {d\lg {T_B} - \frac{{\partial \lg {T_B}}}{{\partial {M_B}}}d{M_B}}  \\
   {d{R_A} - \frac{{\partial {R_A}}}{{\partial {M_A}}}d{M_A}}  \\
   {d{R_B} - \frac{{\partial {R_B}}}{{\partial {M_B}}}d{M_B}}  \\
\end{array}} \right] \\;
 \end{array}
\end{eqnarray}%
where all derivatives are in independent variable set ($\rm{M}$, $\rm{C_{OV}}$, $\rm{X}$, $\rm{Z}$, $\rm{t}$) . All derivatives can be worked out numerically by comparing the stellar model with corresponding ($\rm{M}$, $\rm{C_{OV}}$, $\rm{X}$, $\rm{Z}$, $\rm{t}$) with the stellar models with small variations on them alternately.

Equation (\ref{Appx-relationD}) shows the linear relations between ($\rm{dC_{OV}}$, $\rm{dX}$, $\rm{dZ}$, $\rm{dt}$) and ($\rm{dlgT_{A}}$, $\rm{dlgT_{B}}$, $\rm{dR_{A}}$, $\rm{dR_{B}}$, $\rm{dM_{A}}$, $\rm{dM_{B}}$). The derivatives of ($\rm{C_{OV}}$, $\rm{X}$, $\rm{Z}$, $\rm{t}$) with respect to ($\rm{lgT_{A}}$, $\rm{lgT_{B}}$, $\rm{R_{A}}$, $\rm{R_{B}}$, $\rm{M_{A}}$, $\rm{M_{B}}$) (also the independent variable set) can be calculated based on Equation (\ref{Appx-relationD}). For example, $\rm{\partial (C_{OV},X,Z,t)/\partial R_A}$ can be worked out by setting ($\rm{dlgT_{A}}$, $\rm{dlgT_{B}}$, $\rm{dR_{A}}$, $\rm{dR_{B}}$, $\rm{dM_{A}}$, $\rm{dM_{B}}$) $\rm{=(0,0,1,0,0,0)}$ in the r.h.s. and then the final result vector of the r.h.s. is $\rm{\partial (C_{OV},X,Z,t)/\partial R_A}$.

When the variables $\rm{(x_i)}$ are independent with each other, the standard errors of their functions $\rm{y_j=y_j(x_i)}$ based on the assumption of Gaussian distribution are as follow:
\begin{eqnarray} \label{Appx-Err}
{\sigma ^2}({y_i}) = \sum\limits_j {{{(\frac{{\partial {y_i}}}{{\partial {x_j}}})}^2}{\sigma ^2}({x_j})}.
\end{eqnarray}%

However, in our case, the effective temperatures of the two components of an eclipsing binary star are highly dependent, and the ratio $\rm{T_A/T_B}$ is more accurate \citep{cl07}. This means the restriction $\rm{dlgT_A \approx dlgT_B}$. In this case, we define $\rm{dlgT=dlgT_A=dlgT_B}$ in Equation (\ref{Appx-relationD}) and calculate the derivatives of ($\rm{C_{OV}}$, $\rm{X}$, $\rm{Z}$, $\rm{t}$) with respect to ($\rm{lgT}$, $\rm{R_{A}}$, $\rm{R_{B}}$, $\rm{M_{A}}$, $\rm{M_{B}}$) based on that equation. ($\rm{lgT}$, $\rm{R_{A}}$, $\rm{R_{B}}$, $\rm{M_{A}}$, $\rm{M_{B}}$) are assumed to be independent, thus the standard errors of ($\rm{C_{OV}}$, $\rm{X}$, $\rm{Z}$, $\rm{t}$) can be worked out by using Equation (\ref{Appx-Err}) where $\rm{\sigma lgT}$ is calculated as $\rm{\sigma lgT =(\sigma lgT_A +\sigma lgT_B)/2}$.

\end{document}